\documentclass[twocolumn,secnumarabic,amssymb, nobibnotes, aps, prd]{revtex4}
\usepackage{graphicx}

\begin{document}
\title{Multiple and single measurements of a mesoscopic quantum system with two permitted states}
\author{A.A. Burlakov, V.L. Gurtovoi, S.V. Dubonos, A.V. Nikulov and V.A. Tulin}
\affiliation{Institute of Microelectronics Technology and High Purity Materials, Russian Academy of Sciences, 142432 Chernogolovka, Moscow District, RUSSIA.} 
\begin{abstract} Mesoscopic loop is proposed in many works as possible solid-state quantum bit, i.e. two-state quantum system. The quantum oscillations of resistance and of rectified voltage observed on asymmetric superconducting loops give evidence of the two states at magnetic flux divisible by half of the flux quantum. But our measurements of quantum oscillations of the critical current of these loops have given results coming into irreconcilable contradictions with result of the observations of the quantum oscillations of resistance.
 \end{abstract}

\maketitle

\narrowtext

\section*{Introduction}

One of the most intriguing problems of mesoscopic physics is possibility of quantum superposition of macroscopic states. It  is especially urgent because of the aspiration for realization of the idea of the quantum computation [1]. Ambiguous experiments [2] can not be considered as an evidence of the macroscopic quantum superposition and possibility of solid-state qubit because of the obvious contradiction between quantum mechanics and macroscopic realism [3]. Quantum superposition presupposes that single measurement should give result corresponding to one of the permitted states whereas multiple measurement should give a result corresponding to an average value. We present experimental results corresponding such measurements.
\begin{figure}
\includegraphics{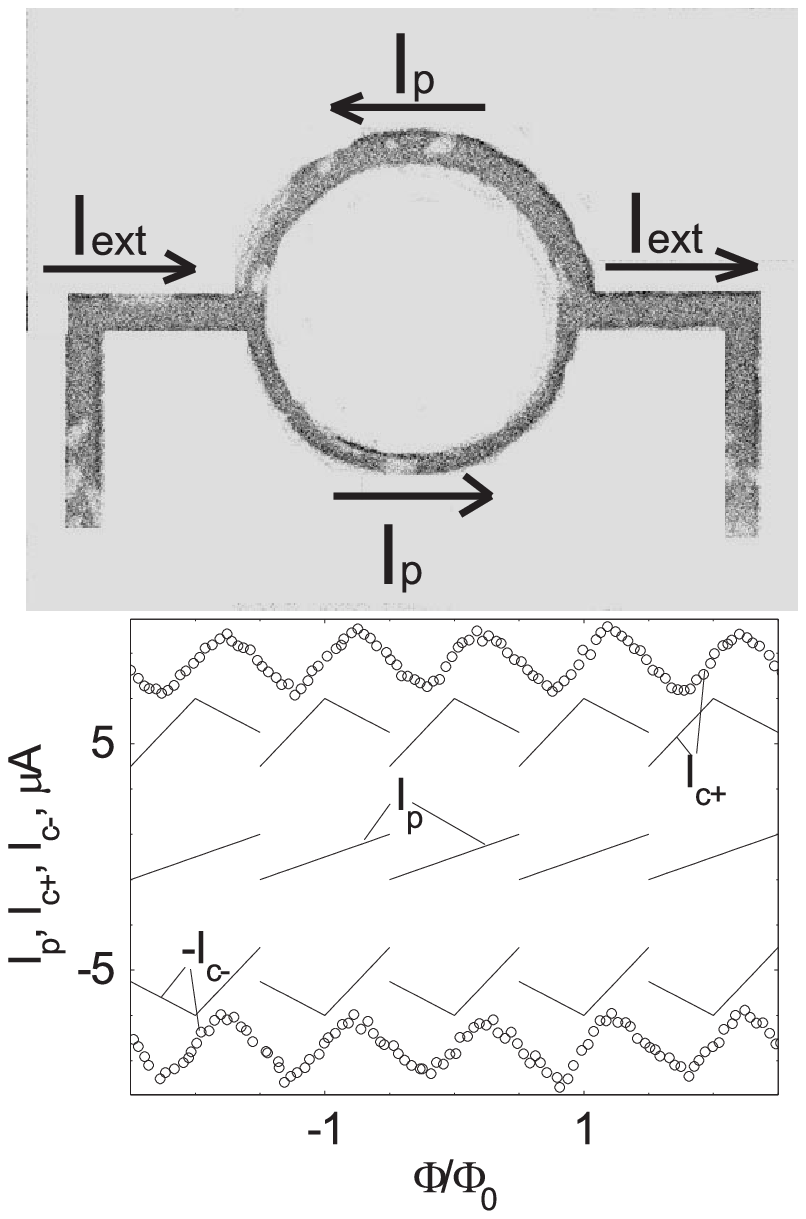}
\caption{\label{fig:epsart} Photo of the asymmetric Al round loop (ring) with semi-ring width $w_{n} = 200 \ nm$, $w_{w} = 400 \ nm$ and magnetic dependencies of the critical current $I_{c+}, I_{c-}$ expected be observed on such ring at $I_{p} = 2 \ \mu A \ (n -\Phi/\Phi_{0})$ and $j_{c}(s_{n}+s_{w}) = 7 \ \mu A$. The experimental dependencies $I_{c+}(\Phi/\Phi_{0}), I_{c-}(\Phi/\Phi_{0})$ measured on this ring at $T = 1.225 K = 0.99T_{c}$ are shown also.}
\end{figure}

\section {Superconducting loop as possible quantum bit}
Superconducting loop interrupted by one or several Josephson junctions is proposed in many publications [4] as possible quantum bit, i.e. a two-state quantum system which can be used as main element of quantum computer. This proposal is based on the assumption on two permitted state in such loop with half quantum $\Phi_{0}= \pi \hbar/e$ of magnetic 
flux $\Phi = (n+0.5)\Phi_{0}$. The authors [4] do not doubt that the two states exist and any observation will find a value corresponding to one of them. But nobody must be sure of anything in the quantum world till unambiguous experimental evidence. A. Einstein, B. Podolsky, and N. Rosen were sure [5] that a process of measurement carried out on a one system can not affect other system in any way. But experimental results [6] have shown that it can and this phenomenon is called now Einstein - Podolsky- Rosen correlation.
\begin{figure}
\includegraphics{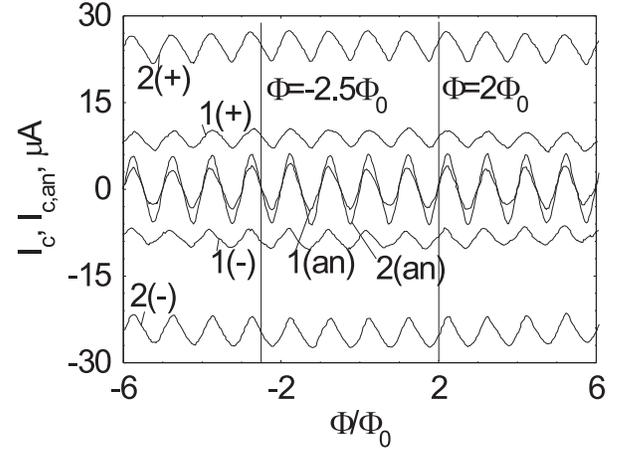}
\caption{\label{fig:epsart} The quantum oscillations of the critical current $I_{c+}$, $I_{c-}$ measured in opposite directions and its anisotropy (an) $I_{c,an} = I_{c+}- I_{c-}$ observed on single loops with $w_{n} = 0.2 \ \mu m$,  $w_{w} = 0.3 \ \mu m$ and $T_{c} = 1.23 \ K$ at (1) $T = 1.271 \ K$ and (2) $1.243 \ K$.}
\end{figure}

The numerous observation of the Little-Parks oscillations [7] of resistance $\Delta R(\Phi/\Phi_{0})$ of superconducting loop [8] prove quantization of velocity circulation $\oint_{l} dl v = (2\pi \hbar/m) (n -\Phi/\Phi_{0})$ of superconducting pairs and that the permitted state with minimum energy has overwhelming probability even at $T \approx T_{c}$. The maximum $\Delta R(\Phi/\Phi_{0}) \propto \overline{v^{2}}$ [9] and zero value of the rectified voltage, corresponding to the average velocity $V_{dc}(\Phi/\Phi_{0}) \propto \overline{v}$ [10], observed at $\Phi = (n+0.5)\Phi_{0}$ may be considered as an experimental evidence of two permitted states with the same minimum energy $\propto v^{2} \propto  (n -\Phi/\Phi_{0})^{2} = (1/2)^{2}$ and $(-1/2)^{2}$: $\overline{v^{2}} \propto (1/2)^{2}+ (-1/2)^{2} = 1/2$ whereas $\overline{v} \propto (1/2)+ (-1/2) = 0$. But it is needed to verify that a single measurement gives a result corresponding $v \propto 1/2$ or $v \propto -1/2 $ state.

\section {Multiple and single measurements of the persistent current}
The measurement of the critical current of an asymmetric superconducting loop, Fig.1, can be used for this verification. The current density equal $j_{n} = I_{ext}/(s_{n}+s_{w}) \pm I_{p}/s_{n}$, $j_{w} = I_{ext}/(s_{n}+s_{w}) \mp I_{p}/s_{w}$ in the loop halves because of the velocity quantization should mount the critical value, $j_{n} = j_{c}$ or  $j_{w} = j_{c}$, at the external current  $I_{c+}, I_{c-} = |I_{ext}| =  j_{c}(s_{n}+s_{w}) - |I_{p}|(s_{n}+s_{w})/s_{n}$ or $I_{c+}, I_{c-} = |I_{ext}| =  j_{c}(s_{n}+s_{w}) - |I_{p}|(s_{n}+s_{w})/s_{w}$ depending on the directions of the external current $I_{ext}$ and the persistent current $I_{p} = 2en_{s}v = 2en_{s}[2s_{n}s_{w}/(s_{n}+s_{w})](2\pi \hbar/ml) (n -\Phi/\Phi_{0})$, Fig.1. One may expect to determine not only value but also direction of the persistent current $I_{p} = (I_{c-} - I_{c+})(s_{w}/s_{n} - s_{n}/s_{w})$ using the critical current $I_{c+}, I_{c-}$ values measured in opposite directions of asymmetric loop with unequal half sections $s_{n}< s_{w}$, Fig.1.

Our measurements of aluminum rings with radius $r = 2 \mu m$, thickness $d = 40-70 \ nm$, semi-ring width $w_{n} = 200 \ nm$, $w_{w} = 400; 300; 250 \ nm$ and systems of such rings at $T < 0.99 T_{c}$ have shown that the whole structure jumps from superconducting to normal state at $I = I_{c+}$ or $I_{c-}$. This means that the ring should remain in superconducting state with the same quantum number $n$ right up to the transition into the normal state and the measurement of $I = I_{c+}$ or $I_{c-}$ should correspond single measurement of the quantum state. One should expect gaps in the $I_{c+}(\Phi/\Phi_{0})$, $I_{c-}(\Phi/\Phi_{0})$ dependencies and maximum of $|I_{p}|(\Phi/\Phi_{0}) \propto |I_{c-} - I_{c+}|(\Phi/\Phi_{0})$ at $\Phi = (n+0.5)\Phi_{0}$, Fig.1.

\begin{figure}
\includegraphics{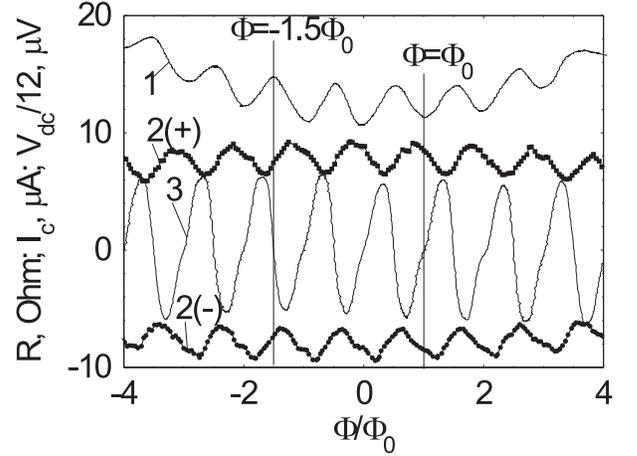}
\caption{\label{fig:epsart} Magnetic dependencies of (1) the resistance $R - 23 \ \Omega $ at T = 1.228 K of (2) the critical current $I_{c+}$, $I_{c-}$ at T = 1.208 K, and (3) the rectified voltage $V_{dc}/12$, induced by the ac current with frequency f = 0.5 kHz and amplitude $I_{0} = 9 \ \mu A$ at T = 1.209 K of 20 loops with $w_{n} = 0.2 \ \mu m$,  $w_{w} = 0.4 \ \mu m$ and $T_{c} = 1.23 \ K$}
\end{figure}

We have obtained identical $I_{c+}(\Phi/\Phi_{0})$, $I_{c-}(\Phi/\Phi_{0})$ dependencies at measurements of four single rings and two systems of identical 20 rings at different temperature, Fig.2,3, which differ from the expected one, Fig.1, in essence. The magnetic dependencies of the anisotropy of the critical current $I_{c,an} = I_{c+}- I_{c-}$, which should be proportional to the persistent current $I_{c,an}(\Phi/\Phi_{0}) \propto - I_{p}(\Phi/\Phi_{0}) \propto -(n -\Phi/\Phi_{0})$, cross zero at $\Phi = n\Phi_{0}$ and $\Phi = (n+0.5)\Phi_{0}$, Fig.2, as well as the one of the  rectified voltage $V_{dc}(\Phi/\Phi_{0}) \propto \overline{n} -\Phi/\Phi_{0}$, Fig.3, corresponding multiple, but not single, measurement of the persistent current states.

It is more strange that the magnetic dependencies of the critical current measured in opposite directions are similar $I_{c+}(\Phi/\Phi_{0}) = I_{c-}(\Phi/\Phi_{0}+\Delta \phi )$ and its anisotropy results from a shift $\Delta \phi = 0.5$ of these dependencies one relatively another. It is very strange that minimum of $I_{c+}(\Phi/\Phi_{0})$ and $I_{c-}(\Phi/\Phi_{0})$ is observed at $\Phi = (n+0.25)\Phi_{0}$ and $\Phi = (n+0.75)\Phi_{0}$ but not at $\Phi = (n+0.5)\Phi_{0}$ as it should be expected and as it is observed in symmetrical ring [11] since the maximum of the Little-Parks oscillations of asymmetric rings is observed at $\Phi = (n+0.5)\Phi_{0}$. We hope that future investigations can clear a nature of this contradictions between results of measurements of $I_{c+}(\Phi/\Phi_{0})$, $I_{c-}(\Phi/\Phi_{0})$ and $\Delta R(\Phi/\Phi_{0})$.

\section {Acknowledgement}
This work has been supported by a grant "Quantum bit on base of micro- and nano-structures with metal conductivity" of the Program "Technology Basis of New Computing Methods" of ITCS department of RAS, a grant of the Program "Low-Dimensional Quantum Structures" of the Presidium of Russian Academy of Sciences and a grant 04-02-17068 of the  Russian Foundation of Basic Research.

\end{document}